\UseRawInputEncoding

{

\documentclass[pre,aps,10pt,onecolumn,twoside,floats,preprintnumbers,floatfix,superscriptaddress,a4paper,showpacs,showkeys,longbibliography]%
{revtex4-2}

\usepackage{amsmath}
\usepackage{amsfonts}
\usepackage{amssymb}

\usepackage{hyperref}

\usepackage{times}
\usepackage{time}
\usepackage{graphicx}

\usepackage{color}

\newcommand{\beq}{\begin{equation}}
\newcommand{\eeq}{\end{equation}}

\def\ie{\emph{i.e.}~}
\def\cf{\emph{cf.}~}

\def\fig#1{\ref{Fig:#1}}
\def\Fig#1{Fig.~\fig{#1}}

\def\eq#1{(\ref{Eq:#1})}
\def\Eq#1{Eq.~\eq{#1}}

\def\rmd{\ensuremath{\mathrm{d}}}

\def\celsius{\ensuremath{^{\circ}\text{C}}}
\def\degree{\ensuremath{^{\circ}}}

\begin{document}
\title{Universality of breath figures on two-dimensional surfaces: an experimental study}

\author{L. Stricker}
\affiliation{\text{ETH Z\"urich, Department of Materials, Soft and Living Materials, 8093 Zurich, Switzerland}}

\author{F. Grillo}
\affiliation{\text{ETH Z\"urich, Department of Materials, Soft Materials and Interfaces, 8093 Zurich, Switzerland}}

\author{E. A. Marquez}
\affiliation{\text{ETH Z\"urich, Department of Materials, Soft Materials and Interfaces, 8093 Zurich, Switzerland}}

\author{G. Panzarasa}
\affiliation{\text{ETH Z\"urich, Department of Materials, Soft and Living Materials, 8093 Zurich, Switzerland}}

\author{K. Smith-Mannschott}
\affiliation{\text{ETH Z\"urich, Department of Materials, Soft and Living Materials, 8093 Zurich, Switzerland}}

\author{J. Vollmer}
\affiliation{\text{University of Leipzig, Institute of Theoretical Physics,  04103 Leipzig, Germany}}



\begin{abstract}
Droplet condensation on surfaces produces patterns, called breath figures. Their evolution into self-similar
structures is a classical example of self-organization. It is described by a scaling theory with scaling functions
whose universality has recently been challenged by numerical work. Here, we provide a thorough experimental
testing, where we inspect substrates with vastly different chemical properties, stiffness, and condensation rates.
We critically survey the size distributions, and the related time-asymptotic scaling of droplet number and surface coverage.
In the time-asymptotic regime they admit a data collapse: the data for all substrates and condensation rates lie on universal
scaling functions.
\end{abstract}

\maketitle
Breath figures are droplets patterns formed by a supersaturated vapour flux
condensing on a substrate \cite{ray11}. They appear in nature, for example when dew
deposits on leaves, spider nets and vegetable fibers. They also have an appealing potential
for technological purposes. Possible applications include dew harvesting devices
for water collection \cite{nik96,clu08,lek11}, heat exchangers with increased efficiency
\cite{ros02,lea06,sik11,bar18}, and patterned surfaces production \cite{bok04,hau04,wan07,ryk11,fen13}.
Breath figure self-assembly has been exploited to fabricate porous bead-on-string fibers
\cite{fen13}. Droplets have been used as a template to produce ordered porous materials for
membrane manufacturing \cite{zha15}, as well as to introduce desired materials inside textiles
by means of three-dimensional porous microstructures \cite{gon17}. Recent studies have shown
that droplet patterns on surfaces can also give origin to structural colours \cite{goo19}.
In all these applications, understanding the droplet formation process and the evolution of
the condensation patterns is a crucial step towards controllability and further technological
development.

The theory of breath figures is based on scaling arguments
\cite{vio88,fam88,fam89,fam89b,mea91,mea92}. The condensation process leading to the
formation of a droplet pattern develops in several phases \cite{bey91}, corresponding to
different time and length scales characterizing the phenomenon. A first nucleation of droplets
(primary nucleation) is followed by their initial growth as a monodisperse population.
After some time, the droplets start to merge, releasing space on the substrate, which is used for
further nucleation (secondary nucleation). The distribution continues to evolve, becoming polydisperse
and eventually self-similar \cite{vio88,fam88,kol89,bri98}. Scaling concepts, closely related to fractals
theory, can be used to describe the evolution of the droplet distribution~\cite{bri98,bla12}.
Experiments \cite{bey86,vio88,car97,had98,bla12} and simulations \cite{fam89,ulr04,bla12,str15} have shown
that, in the late-time regime, the droplet size distribution is bimodal, with two well-separated parts.
In particular, it features a bell-shaped peak, corresponding to the monodisperse population of the largest
hence oldest droplets, and a power law distribution of smaller droplets, which is terminated by a cutoff
function at the nucleation length scale. Scaling manifests itself in a data collapse of droplet
distributions taken at different times, and in the time-dependence of the moments of the distribution.
In the long-time regime they approach power laws with exponents that can be expressed in terms of a single
non-trivial exponent, denoted as ``polydispersity exponent $\tau$.'' In particular, the asymptotic
time decay of the droplet number and the porosity (ratio between non-wetted area and total substrate area)
are described by the same exponent.
It was widely expected that the polydispersity exponent is a universal number, depending only on the dimensionality
of the system \cite{vio88,fam88,fam89,fam89b,mea89,mea91,mea92,bla00}. Its value was calculated \cite{bla00}
by assuming universality \cite{smo16}.
However, the exponents found in recent numerical simulations \cite{bla12,str15} differ clearly
from the prediction. This calls for experimental studies.
So far experimental studies have mostly addressed the early stages of the droplet nucleation \cite{bey86,bey91,fri91},
and the initial phases of the polydisperse transient regime \cite{bri91,gua13}, with noticeable exceptions in~\cite{bla12,bar18}.
The impact of surface properties and condensation rates has not been examined.

Here, we present extensive experimental data for breath figures on a range of substrates with different stiffness,
surface chemistry, and temperature. The different regimes of surface coverage are discussed with an emphasis on the self-similar phase.
We observe the predicted scaling and critically survey the predicted data collapse
for all data of all experimental settings. The conclusion, that there are universal scaling functions, is substantiated by Kinetic Monte Carlo simulations.

\paragraph*{Evolution of droplet patterns. ---}
We induce the nucleation and growth of water droplets on substrates consisting of a glass cover slip either coated
with a $30\;\mu$m layer of silicone (Dow Corning Sylgard 184) or silanized with tridecafluoro--1,1,2,2--tetrahydrooctyl
trichlorosilane (SiHCl$_3$) or hexametildisilazane (HMDS). The upper side of the substrates is exposed to a steady flux of air saturated with water at room temperature.
On the bottom, they are in contact with a temperature-controlled plate at a set-point temperature $T_p^*$ of 5$\celsius$.
We image the droplets from the top with a dissecting microscope (Nikon SMZ80N). The smallest measured droplets have a radius of about $15\;\mu$m,
the largest about $2.5\:$mm.
In each image, we identify the center and radius of each droplet.
We acquire images of the droplets over logarithmically spaced time intervals between $0.1\:$sec and $10\:$h, from the moment the first visually resolvable droplets appear.
Using silicone substrates allows us to reduce the stiffness, and hence the number of nucleation sites respect to glass \cite{pha17}. The static contact angle $\theta_c$ is measured
via side imaging, with a CMOS camera (Thorlabs, DCC3240M) and LED back-illumination, with a precision of $2\degree$. On all substrates droplets can be considered
hemispherical ($\theta_c \sim 90\degree$) except on HMDS-glass ($\theta_c = 67\degree$). Full details on the experimental setup, data acquisition, and analysis are given in
Sect.~\ref{Sec:setup}--\ref{Sec:imaging} of the Supplementary~Material.

\begin{figure}
\centering
\includegraphics[width=0.96\columnwidth]{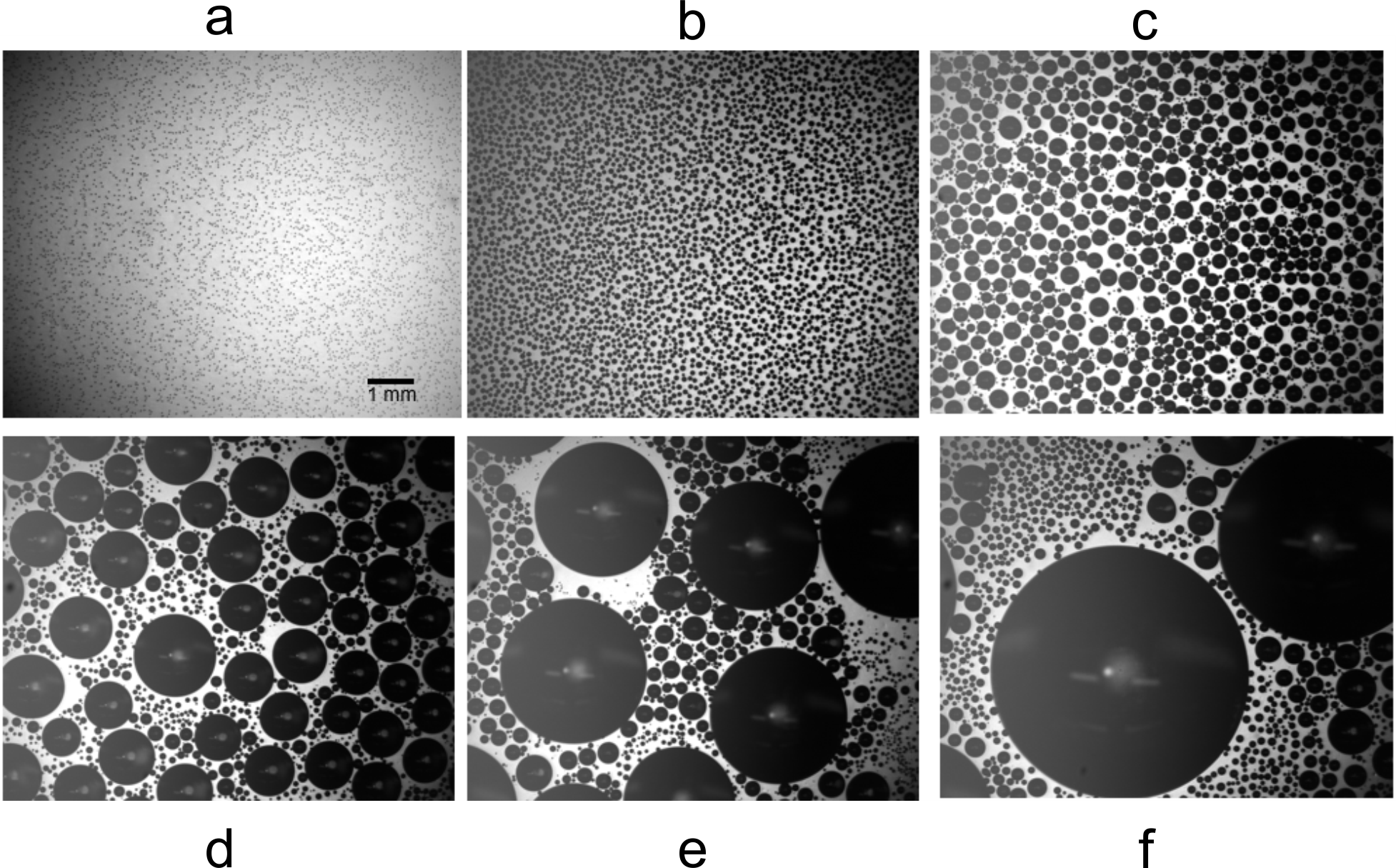}
\caption{Time evolution of the droplets condensation pattern on a silicone substrate
 with elastic modulus $E = 2\:$MPa. Images are taken (a) 1 sec,  (b) 10 sec, (c) 100 sec, (d) 15 min,
(e) 250 min, and (f) 300 min after the beginning of the condensation process.}
 \label{Fig:patterns}
\end{figure}

Representative snapshots of droplet nucleation, growth, and coalescence on a silicone substrate with elastic modulus
$E = 2\:$MPa  are shown in Fig.~\ref{Fig:patterns}.
After an initial burst of nucleation, (Fig.~\ref{Fig:patterns}a), the droplets grow with roughly uniform size, (Fig.~\ref{Fig:patterns}b).
After about one minute, the droplets start to come into contact  and coalesce (Fig.~\ref{Fig:patterns}c).
New droplets nucleate in the gaps between larger ones, and  the range of droplet sizes grows (Fig.~\ref{Fig:patterns}d-f).

\begin{figure}
\centering
\includegraphics[width=0.98\columnwidth]{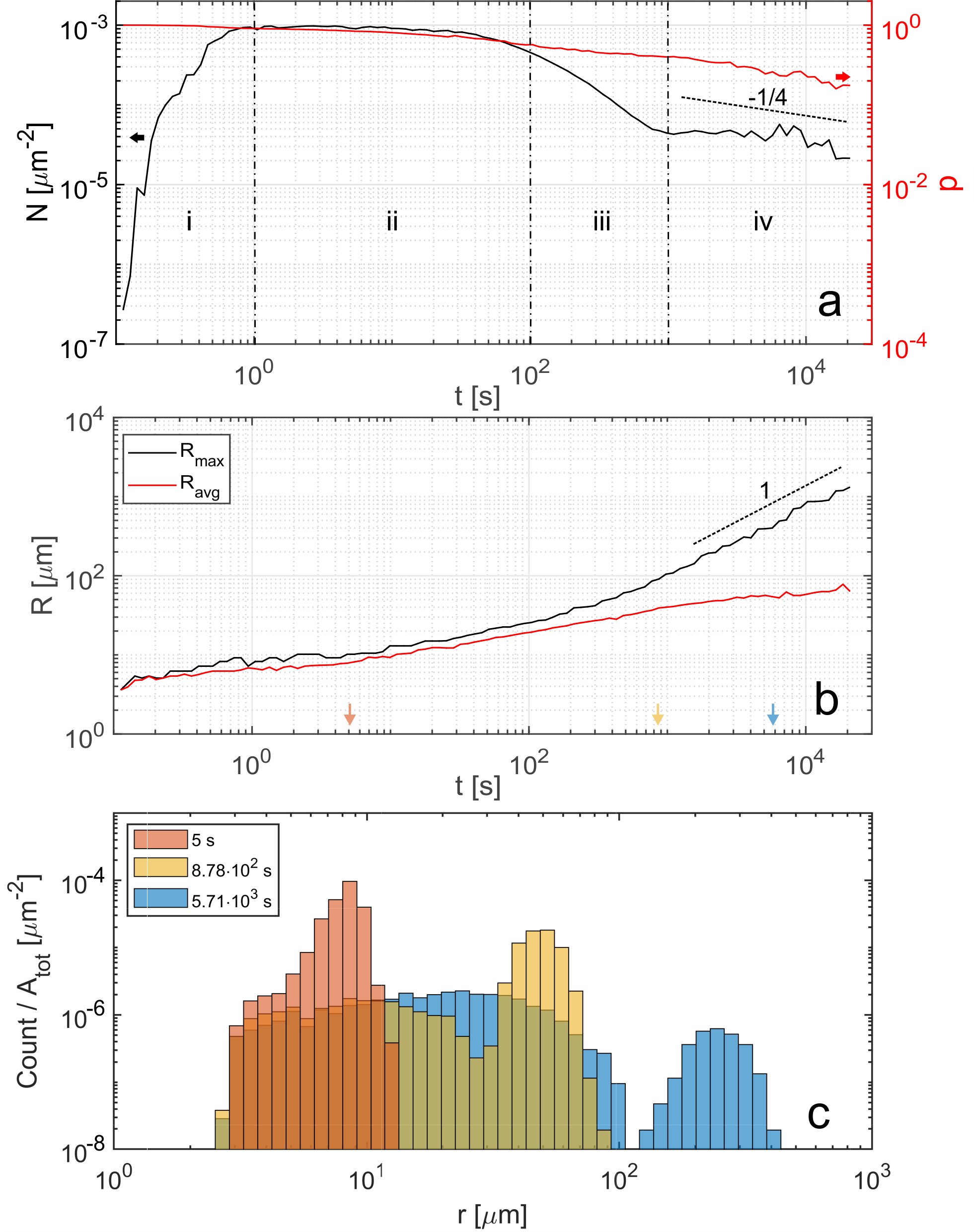}
\caption{Stages of growth. (a) Time evolution of the number of droplets per unit area
(black, left vertical axis) and porosity (grey, red online, right vertical axis).
(b) Time evolution of the maximum (black) and average (grey, red online) radius of a droplets population
condensing on a silicone substrate ($E = 1\:$MPa).
(c) Frequency histograms of the radii $r$ of the droplet population at different times:
$1\:$min (red), $15\:$min (yellow) and $95\:$min (blue)
from the beginning of the condensation process}
 \label{Fig:stages}
\end{figure}

Four stages of the condensation process emerge clearly when we plot the total number of droplets per unit area as a function of time,
$N(t)$, as shown in Fig.~\ref{Fig:stages}a (black line, left vertical axis). The nucleation stage lasts for the first second. It is characterized by a rapid increase in the number of droplets and it is labelled
(i) in Fig.~\ref{Fig:stages}a. In the uniform growth stage, labelled (ii) in Fig.~\ref{Fig:stages}a, the number of droplets remains essentially fixed and the
mean and maximum droplet
radii increase, as shown in Fig.~\ref{Fig:stages}b.
Throughout the nucleation and growth stages, the droplet size distribution remains unimodal,
as shown by the red histogram  in Fig.~\ref{Fig:stages}c.
In the early coalescence stage, labelled (iii),  the number of droplets per unit area steadily decreases (Fig.~\ref{Fig:stages}a),
the droplets growth accelerates, as shown in Fig.~\ref{Fig:stages}b, and the size distribution becomes bimodal, as shown
by the yellow histogram in Fig.~\ref{Fig:stages}c.
In the late coalescence stage, starting after about $10^3$ s, labelled (iv), the number of droplets decays more slowly than before
(Fig.~\ref{Fig:stages}a), while the spread between the maximum and mean droplet radii  widens, as shown in Fig.~\ref{Fig:stages}b, reflecting the broadening of the underlying size distribution (Fig.~\ref{Fig:stages}c, blue histogram).
As the droplets distribution grows, the free area on the substrate decreases. This decay is quantified by the porosity, \beq
p(t) = 1 - \sum_i \frac{A_i(t)}{A_{tot}}\,
\label{Eq:porosity, def}
\eeq
where the index $i$ labels the $i^{th}$ droplet,
$A_i = \pi R_i^2$ is its wetted area, with $R_i$ its time-dependent radius, and $A_{\text{tot}}$ is the total substrate area \footnote{Note that the porosity is an instantaneous measure of the area not covered by droplets at a certain time, thus it differs from the 'visited area' \cite{mar95}, which is the area that has been occupied by droplets at any previous time}. Experimentally, the porosity is calculated from the droplet coordinates and radii.
Prior to the nucleation of droplets, for an empty surface, the porosity is equal to one.
As the area covered by droplets grows in time, the porosity decays, as shown
in \Fig{stages}a (grey line, red online, right vertical axis).

\paragraph*{Scaling of  the droplet number density. ---}
We now evaluate the droplet size distributions. The probability density function $n(s,t)$ represents the number of droplets of size $s$ per substrate unit area, per unit size. The ``size'', proportional to the droplets mass, is defined as $s=r^3$, where $r$ is the radius and $n$ has units of $m^{-5}$. During the late-stage scaling regime, $n$ should adopt a scaling form. In particular, it is expected \cite{fam88,fam89} that the intermediate portion of the distribution, excluding the tails of the smallest and the largest droplets, scales as  $n(s,t) \sim [s/\Sigma(t)]^{-\tau} \Sigma(t)^{-\theta}$, where $\theta$ is a trivial exponent, depending on the dimensionality of the system, $\tau$ is the polydispersity exponent and $\Sigma(t)$ is the maximum droplet size at time $t$. The exponent $\theta$ must take the value 5/3 for 3-dimensional droplets on a 2-dimensional substrate, such that the dimensions of $n$ and the scaling expression match. The exponent $\tau$ is expected to take a value of $ 19/12$ \cite{bla00}.
The resulting power laws are indicated by dashed lines in the plots of our data.
Further discussion is provided in Sect.~\ref{Sec:scaling of size distribution} of the Supplementary Material.
We analyze $n$ both for experiments and simulations. In particular, we perform Kinetic Monte Carlo simulations \cite{gil76,bat08,jan12_BOOK, soe18} on a $1200\times1200$ square lattice with periodic boundary conditions and a constant water flux impinging onto the surface. The simulations account for droplet nucleation and growth as well as for merging events. A full description is provided in Sect.~\ref{Sec:simulations} of the Supplementary~Material.

\begin{figure}
\centering
\includegraphics[width=0.95\columnwidth]{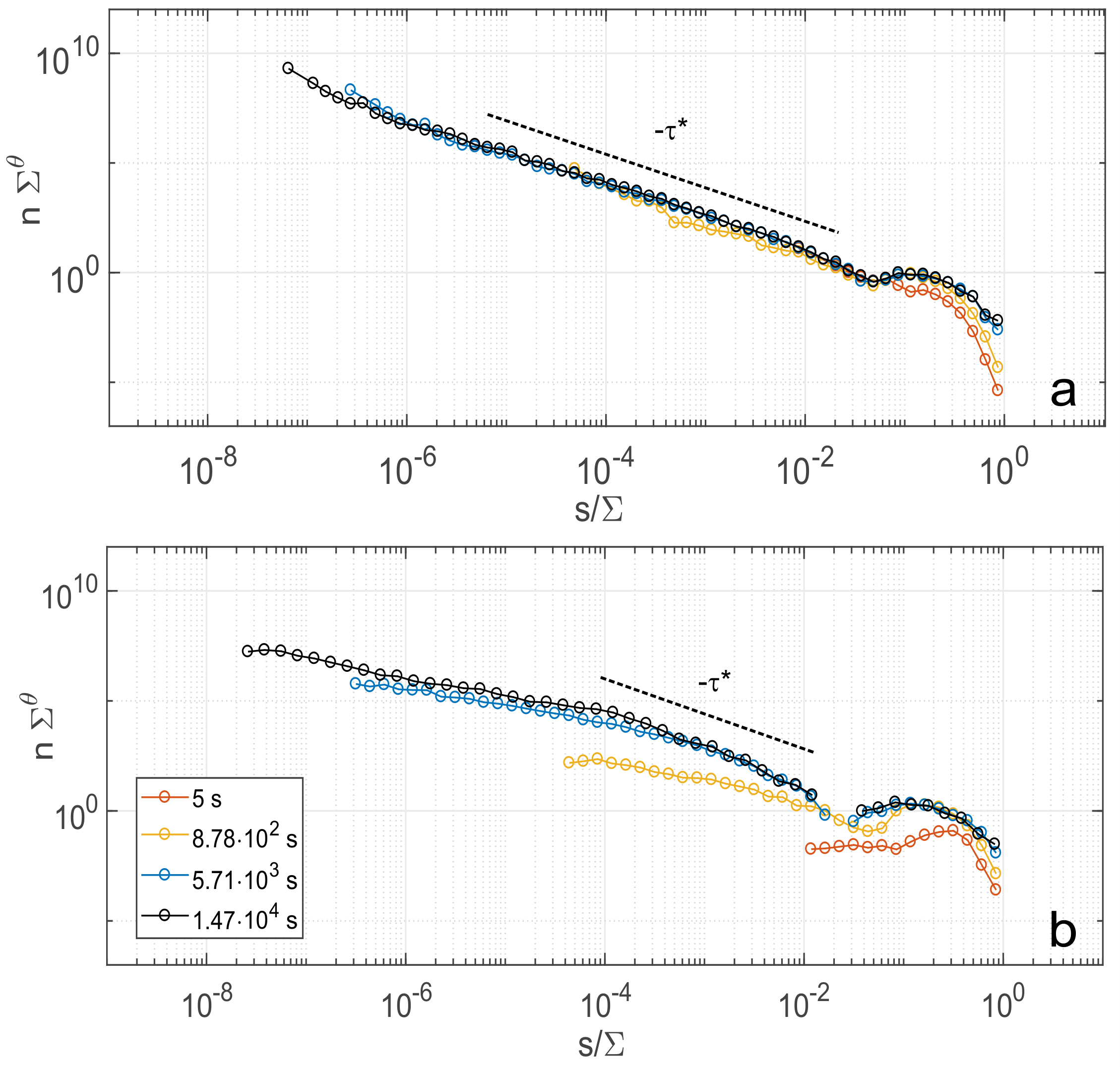}
\caption{Rescaled droplet number density as a function of the size $s / \Sigma$ where $s = r^3$,
$\Sigma = \max{s}$ and $\theta = 5/3$ (a) from simulations and (b) from experiments, at different
times, indicated in the legend.
The dashed lines represent the theoretical prediction for the
polydispersity exponent $\tau^\star = 19/12$.}
\label{Fig:prob density}
\end{figure}

To compare our findings to the theoretical predictions, we plot the rescaled droplet number density, $n(s,t)\Sigma^{\theta}$, in Fig.~\ref{Fig:prob density}. In this plot, the droplet sizes are normalized by the maximum droplet size observed at that time point, $\Sigma(t)$. Thus, large droplet peaks line up at $s/\Sigma\approx1$ for all times.
As time progresses, the small droplet peak becomes broader and has a maximum value close to the smallest resolvable size.

At large times, the distribution presents three distinct features: an intermediate self-similar range where $n \sim (s/\Sigma)^{-\tau}$, a monodisperse bump
describing the large droplets, and a tail describing the small droplets.
Such features clearly appear in both our numerical and experimental results (Fig.~\ref{Fig:prob density}).
For all data the large-droplet cutoff emerges for $s/\Sigma \gtrsim 10^{-2}$.
The simulations show a power law over around four decades, in the intermediate scaling range, for $10^{-6} < s/\Sigma < 10^{-2}$.
In contrast, for the experimental data, the scaling range is limited to at most two decades, $10^{-4} < s/\Sigma < 10^{-2}$, even for our latest time $t=1.47 \; 10^4\,$sec.
The small-droplet cutoff is much broader for the experiments and it covers roughly three decades, with the cross-over at $s/\Sigma \approx 10^{-4}$. The experiments cannot be further extended in time, since gravitational effects impact the droplet shape when the large droplets approach the capillary length, $\sqrt{\gamma/\rho g} \approx 2.6 ~ \mathrm{mm}$.

\paragraph*{Porosity and Droplet Number. ---}
The scaling form of the distribution entails that the porosity and the droplet number exhibit a power-law decay \cite{fam88,fam89}. According to the theory, in the scaling regime, the radius of the largest droplets should scale as $R\sim t ^{\nu}$ with exponent $\nu=1$, and both the number of droplet per unit surface $N$, and the porosity $p$ should evolve as a power law $N \sim p \sim t^{-k}$ \cite{fam88,fam89}, with exponent
\beq
  k = 3\nu (\theta - \tau) \, .
  \label{Eq:compatibility condition}
\eeq
The derivations are provided in Sect.~\ref{Sec:largest droplet size}, \ref{Sec:derivation p(t)} and~\ref{sec:derivation N(t)} of the Supplementary~Material. For $\theta = 5/3$ and Blackman's prediction $\tau = \tau^* = 19/12$ \cite{bla00}, we expect that $k = k^* = 1/4$, where the asterisks denote the specific theoretical values.

\begin{figure}
\centering
\includegraphics[width=0.95\columnwidth]{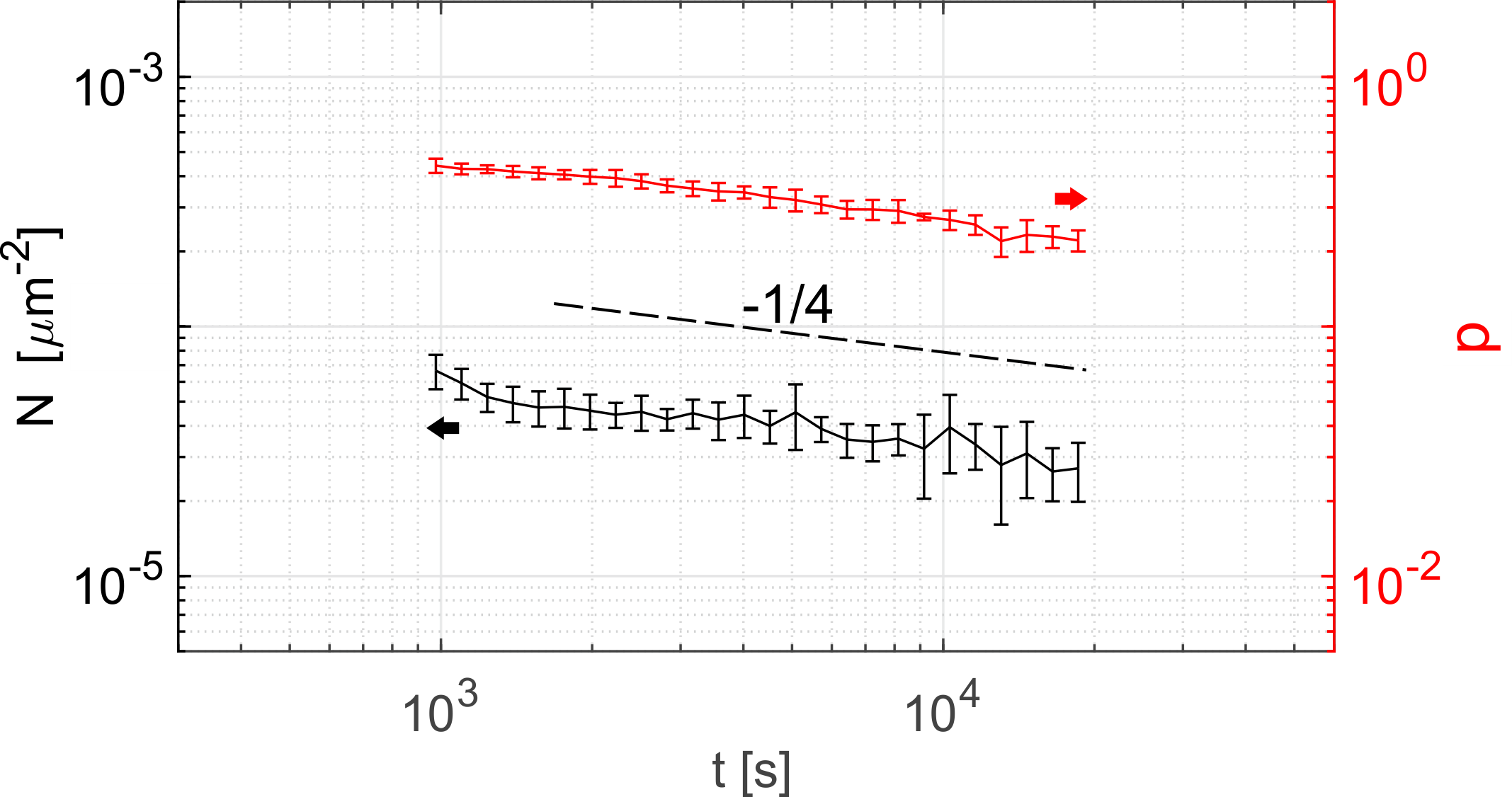}
\caption{Late-time evolution of porosity (grey, red online, right vertical axis) and number of droplets per unit area (black, left vertical axis), averaged over
6 experiments. The error bars represent the standard deviation.}
\label{Fig:slopes}
\end{figure}

The temporal growth of the maximum droplet radius is shown in Fig.~\ref{Fig:stages}b.
In the last time decade of the experiment, it follows the predicted power law with exponent $\nu=1$. \Fig{slopes} shows the average of the porosity (grey, red online, right vertical axis) and number of droplets per unit area (black, left vertical axis) averaged over 6 experiments.
The error bars represent the standard deviation at each instant. Within the error bars, the decay rate of both the porosity
and droplets number are compatible with the theoretical exponent $- 1/4$ (dashed line), but deviations up to 20\% remain possible \cite{bla00,str15}.
Note that the exponent $k$ is extremely sensitive to any variation of the exponent $\tau$. A deviation $\Delta\tau$ from
the theoretical prediction $\tau^*$, would be enhanced by a factor of three in the deviation from $k^*$, thus resulting in a twentyfold increase
of relative deviations, $ \Delta k / k^* =  3 \Delta\tau / (1/4) = 12 \tau^* \Delta\tau/\tau^* = 19  \Delta\tau / \tau^*$. Moreover, the integral quantities of porosity and droplet number are easier to evaluate than the full droplet distribution.

\paragraph*{Condensation for different surfaces and water fluxes. ---}
In view of its robust scaling in the late time regime, we now use the porosity time decay to compare condensation for different surfaces and condensation fluxes. For the porosity the theory predicts that
\begin{align}\label{eq:porosity}
  p(t)
  &= \left( \frac{s_p}{x_p \, \Sigma(t)} \right)^{\theta-\tau}
    \quad\text{with}\quad
    \Sigma(t) = R_{max}^3 = \left( \frac{\pi\Phi}{3\alpha} t \right)^3 \, .
\end{align}
Here, $\Phi$ is the constant water flux, \ie the water volume deposited on the substrate per unit surface
per unit time, $\alpha = (\pi/3)(2+\cos\theta_c)(1-\cos\theta_c)^2/(\sin\theta_c)^3$ is a geometrical factor accounting for the contact angle $\theta_c$,
measured through the liquid phase, $s_p/x_p\Sigma(t)$ is the width of the scale separation between smallest and largest droplets;
the non-dimensional sizes $s_p / \Sigma(t)$ and $x_p$, where $s_p$ and $x_p$ are constants, represent indeed the cutoffs
for small and large droplets respectively. Full details are provided in Sect.~\ref{Sec:derivation p(t)} of the Supplementary Material. For this
discussion we note that, due to the small exponent $\theta-\tau = 1/12$, the ratio $(s_p/x_p)^{\theta-\tau}$ in Eq.~\ref{eq:porosity}
can change by at most $30\%$, even for $s_p/x_p$ differing by two orders of magnitude for two vastly
different materials. Hence, it remains practically unvaried, and the impact of $\Phi/\alpha$
on the time evolution of $\Sigma(t)$ is expected to be the only noticeable parameter influencing the asymptotic evolution of the porosity.

\begin{figure}[h]
\centering
\includegraphics[width=0.95\columnwidth]{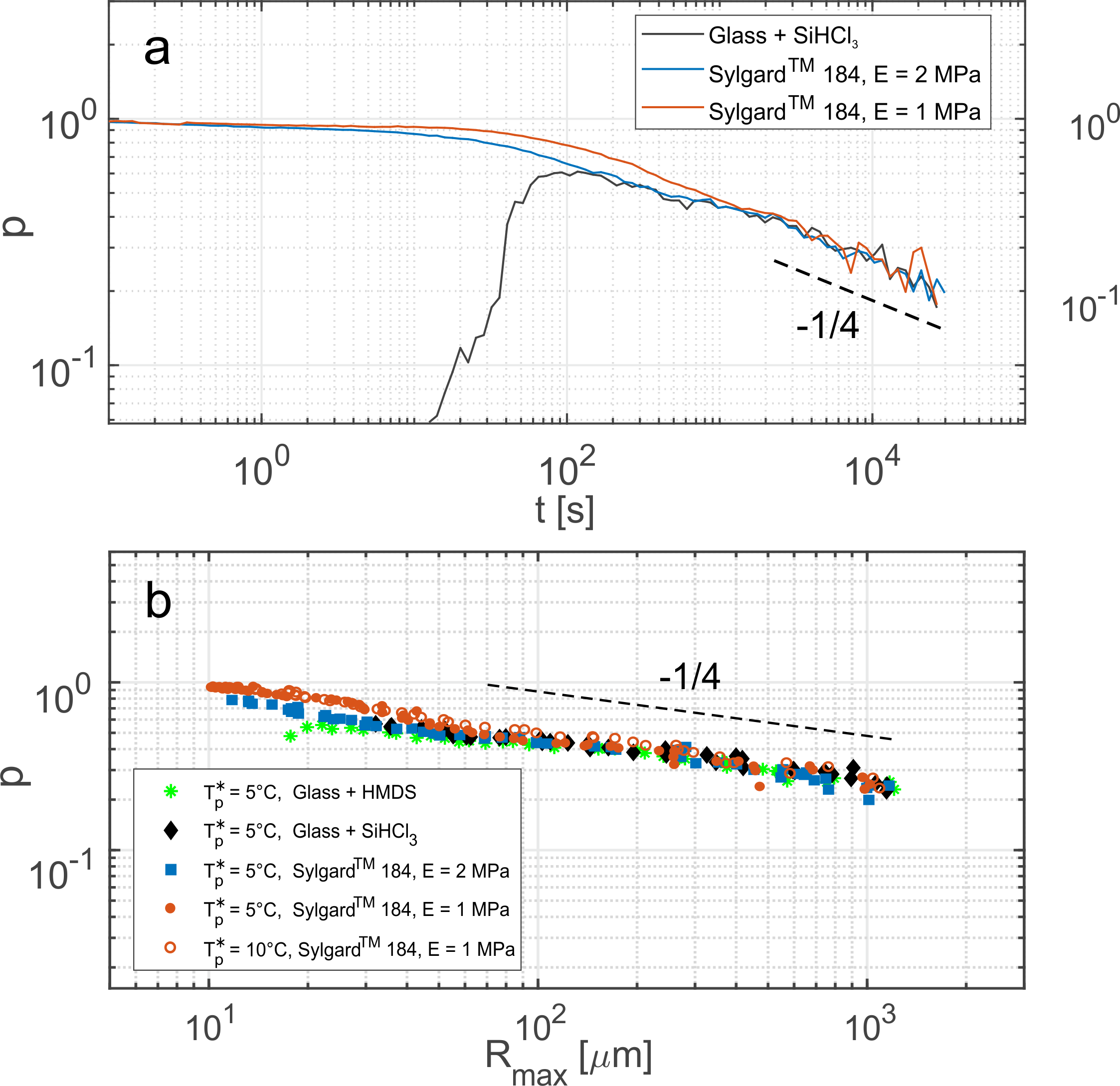}
\caption{(Colour online) Evolution of porosity (a) as a function of time for different rigid surfaces with $T_p^*=5\degree$~C: fluor-silanized glass (black) and silicone with  $E = 2$~MPa (dark grey, blue online)
 and $E = 1$~MPa (light grey, red online) (b) as a function of maximum droplet radius, for different contact angles $\theta_c$
 and vapour fluxes, regulated by changing the plate temperature $T_p^*$.
 The substrates are: HDMS-silanized glass ($\theta_c = 67\degree$) with $T_p^*=5\degree$~C (green asterisks), fluor-silanized glass ($\theta_c = 92.2\degree$) with
 $T_p^*=5\degree$~C (black filled diamonds), 2~MPa silicone substrate ($\theta_c = 94.3\degree$) with $T_p^*=5\degree$~C (blue filled squares), 1~MPa silicone substrate ($\theta_c = 95.5\degree$) with $T_p^*=5\degree$~C (red filled circles), and $T_p^*=10\degree$~C (red empty circles).}
 \label{Fig:porosity diff}
\end{figure}

\Fig{porosity diff}a shows the time evolution of the porosity for three different surfaces with the same temperature of the cooling plate, $T_p^* = 5\celsius$, and the same contact angle up to experimental precision. The light and dark grey curves (red and blue online) correspond to 1 and 2 MPa silicone surfaces, respectively, while the black curve shows the porosity for fluor-silane coated glass. The two silicone surfaces behave similarly, with a monotonic decrease in the porosity. However, the softer surface ($E = 1$~MPa) is populated at a slower rate. Hence the drop in porosity occurs later, as one can observe for times around $100$~s.
The silanized glass surface behaves very differently in the early stages of condensation.
It has a very high nucleation rate and is covered by tiny droplets almost immediately, such that the porosity drops to a very small value. Initially, the droplets are so small and so densely packed that they cannot be individually resolved (Fig.~\ref{Fig:glass, nucleation} in Supplementary Material). Between $10$~s and $100$~s the droplets start to
merge such that the areas in between can be discerned, and the porosity rises towards the values observed for the silicone surfaces. In the late-time scaling regime the surface is entirely covered by water droplets, and the flux $\Phi$ onto the droplets is solely determined by the cooling plate temperature, such that Eq.~\ref{eq:porosity} predicts the same power law for the three systems.
Despite dramatic difference in initial droplet nucleation and growth, at late times, $t \gtrsim 1000$~s, all data fall exactly on top of each other. The three surfaces do not only share the same scaling exponent, but also the same prefactor of the power law.

In \Fig{porosity diff}b we compare condensation on four different types of substrates with different stiffness and contact angle $\theta_c$, and we also vary the vapour flux $\Phi$ by
changing the temperature of the cooling plate $T_p^*$. The porosity of all data falls on a
single scaling function when plotted as function of the maximum droplet radius. We interpret this very good, parameter-free data collapse as strong evidence for universality.

\paragraph*{Conclusion. ---}
We present a series of experiments and simulations where a time-constant uniform water vapour flux condenses on rigid cold surfaces.
The emerging droplets patterns undergo four stages on their way to organize into a self-similar arrangement whose number densities feature non-equilibrium scaling (see~Fig.~\ref{Fig:stages}):
(i) a first wave of nucleation of droplets; (ii) uniform growth of roughly equally spaced and monodisperse droplets;
(iii) early coalescence, releasing surface area formerly occupied by the first generation of droplets; and (iv) re-population of the gaps between droplets and emergence of a self-similar droplet pattern. In the self-similar regime the droplet number densities at different times admit a data collapse, \Fig{prob density}. The scaling of the number densities implies a power law decay of the droplet number and the porosity, \ie the area not covered by droplets.
We showed here that substrates with vastly different surface properties evolve towards \emph{identical} power laws with matching exponents and pre-factors, where different surface fluxes are fully accounted for by adopting the maximum droplet radius as a time variable
(\cf~Eq.~\eqref{eq:porosity}). These findings provide compelling evidence for universal scaling of the asymptotic self-similar regime of breath figures.

\begin{acknowledgements}
We would like to express our gratitude to Eric Dufresne, in whose lab the present experiments have been performed, for his input and valuable suggestions on how to improve earlier versions of the manuscript.
We are thankful to Rob Style, for his help with the design of the experimental setup, and to Kathryn Rosowski and Qin Xu for experimental support. We acknowledge Nicolas Bain, Daniel Dernbach and Robert Haase for insightful discussions, and we thank Martin Callies for feedback on the manuscript.

LS designed, performed and analyzed the experiments. FG and EAM performed the simulations. GP and KSM contributed to the design of experiments and the realization of the samples. LS and JV developed the theory and wrote the manuscript. All the authors provided feedback on the manuscript.
\end{acknowledgements}



%


\clearpage
\section*{Supplemental Material}
\label{Sec:materials and methods}

\subsection{Experimental setup}
\label{Sec:setup}
\begin{figure}
\includegraphics[width=0.75\textwidth]{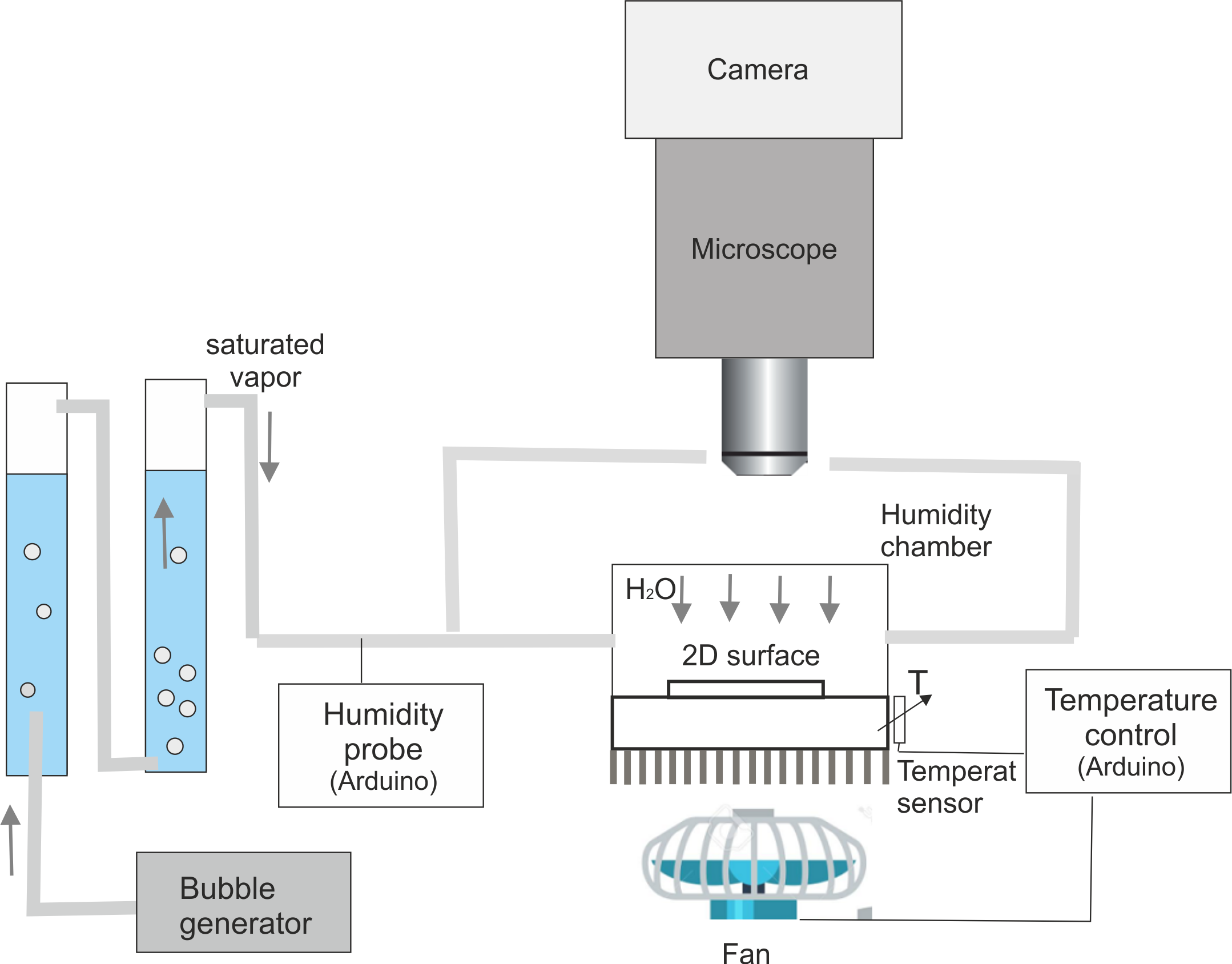}
\caption{Sketch of the experimental setup}
 \label{Fig:setup}
\end{figure}

A sketch of the experimental setup is displayed in \Fig{setup}.
We induce the nucleation and growth of water droplets on a rigid substrate (sample),
inside a condensation chamber (6$\times$5$\times$4~cm).
The sample is placed in direct contact with a 4-mm thick aluminium plate-fin
6$\times$5~cm. The plate is positioned on top of a 12~V fan, 
 and acts like a heat sink.
The temperature control of the condensation substrate is realized by controlling the temperature
of the aluminium plate with a computer-regulated feedback mechanism.
To this aim, a thermistor (NTC 30~k$\Omega$, Amphenol Advanced Sensors) measures the temperature
of the plate and sends the signal to a PID controller, programmed on an Arduino microcontroller.
Both the thermistor and the fan are connected to the PID controller.
The PID controller compares the measured temperature of the plate, $T_{p}$, with the desired
temperature, $T_p^*$, and varies the speed of the fan, until $T_{p}=T_p^*\pm0.1\celsius$.
To guarantee a constant water deposition rate, the sample is placed inside a humidity chamber.
A constant flux of saturated vapour is introduced through two openings on opposite sides of the chamber.
To generate such a flux, a pump (Tetra APS50 Air Pump) insufflates
a constant air flux at the bottom of a deionized water column (5~cm diameter $\times$ 50 cm height).
The air bubbles raise through the column collecting saturated vapour. From the top of the column,
they are conveyed into a second column, repeating the process. The humidity of the vapour flux at the exit of the second
column is measured with a humidity sensor (393 - AM2302 Digital Temperature and Humidity Sensor, 5V, Adafruit),
connected to a second Arduino microcontroller. The vapour flux is turned on before lowering the temperature of
the plate. Within few seconds, the humidity increases up to the point where it reaches full saturation (100\%) and the measure
(which has a precision higher than 0.01\%) does not change anymore.
All our experiments are done in full saturation conditions of the air flux (100\% humidity), with a
constant temperature difference between the controlled lab environment (24\celsius) and the aluminium
plate (5\celsius~or 10\celsius). We choose to change the plate temperature to vary the deposited water flux,
as this provides the best experimental controllability and reproducibility, respect to changing the flow or the thickness of the substrate.
The samples are imaged from the top with a Nikon SMZ80N microscope.
Recording is done with a ThorLabs USB 3.0 Digital Camera. To improve the imaging, a LED light (ThorLabs MCWHL5) is
shed through one of the objectives of the microscope and a 100~$\mu$l water droplet is deposited between
the sample and the aluminium plate.

\subsection{Sample fabrication}
\label{Sec:samples}
The condensation experiment is performed with substrates made of two different materials:
silicon gel and hydrophobic glass. 
The silicone gel substrates are fabricated by spin-coating a layer of polydimethylsiloxane (PDMS)
on top of a glass cover slip (Menzel-Gl\"{a}ser 24x50~mm, N.~1.5).
For the PDMS preparation, we use a commercial binary mixture of Sylgard\texttrademark 184 Silicone Elastomer (Dow Corning,
Base and Curing Agent). The two components are mixed with different mass ratios (4:1 and 10:1) to achieve
different stiffness of the condensation substrates. The mixture is degassed in vacuum and
spin-coated on top of the cover-slip for 1 min at 1000~rpm. The samples are then cured at 40\celsius, for 7 days.
The stiffness of the cured substrate is measured through the Young's elastic modulus $E$,
by means of a compression test performed with a texture analyzer (Stable Microsystem TA.XT plus).
The 4:1 and 10:1 substrates have an elastic modulus of $E = 2\pm 0.18$~MPa and $E = 1 \pm0.2$~MPa respectively.
The glass substrates are made hydrophobic by silane vapour deposition, using the following procedure:
a glass cover slip (Menzel-Gl\"{a}ser 24x50~mm N.~1.5) is exposed to UV radiation for 20 minutes.
Subsequently, it is positioned inside a vacuum desiccator together with a 1~ml
droplet of tridecafluoro -1,1,2,2 - tetrahydrooctyl trichlorosilane 97\%. We generate a 5 mbar depression inside
the desiccator and we isolate it from the air-void line. After 24~h, we remove the glass from the desiccator, we wash it
with toluene and we vacuum dry it.
We measure the static contact angle of the droplets $\theta_c$ on the different substrates by means of side imaging with a
CMOS Camera (Thorlabs, DCC3240M) and back-illumination with a $3.5''\times 6''$ white LED (Edmundoptics). We find a static
contact angle of the order of 90\degree for silicone and fluor-silanized glass ($95.5\degree\pm 2\degree$ for 4:1 silicone,
$94.3\degree\pm 2\degree$ for 10:1 silicone and $92.2\degree\pm 2\degree$ for fluor-silanized glass), while HMDS-silanized
glass has a contact angle of $67\degree\pm 2\degree$.
Given the slow speed of the condensation process, the droplets can be considered in their equilibrium shape during growth by direct water deposition.
Though such shape may change during droplets merging, the relaxation time of a droplet formed by coalescence is very fast
(below few seconds for the largest droplets in the late regime), hence droplets with a non-circular footprint are rarely captured in the
images and statistically irrelevant for the analysis.

\subsection{Image processing}
\label{Sec:imaging}
The image processing is carried out with a combination of Fiji and Matlab scripts.
In order to eliminate artifacts due to possible uneven illumination, we first subtract the background
from the original images, by using the Fiji plugin 'Pseudo flat field correction' of the Jan Borcher's Biovoxxel
toolbox \cite{bro14_BioVoxxel}.
The images are then converted from grey scale to black and white by means of the 'Enhance local contrast (CLAHE)'
\cite{zui94} or with the standard 'Adjust/Threshold' Fiji plugin. The droplets appear as black areas on a white background.
The binarized images are visually inspected to correct the imperfections due to reflections on the surface of the
large droplets. We use the 'Binary/Fill holes' plugin, to correct the images where white areas appear
in the middle of the large droplets, due to such reflections. The segmentation of connected droplets is done with Michael Schmid's Fiji
'Adjustable Watershed' plugin \cite{sch12_plugin}, with an appropriately selected tolerance parameter and subsequently manually
corrected when required. The detection of the radius and the coordinates of the centre of the droplets is performed with a Matlab
self-developed code. The static contact angle for the different substrates are measured with Stalder's Fiji Low Bond Axisymmetric Drop
Shape Analysis plugin \cite{sta10}.

\subsection{Data collection and statistical analysis}
\label{Sec:statistics}
For the droplet number density, $n(s,t)$, we consider only the droplets whose centre of mass resides inside the field of view.
For each time point, the corresponding droplets size distribution is calculated based on at least 3000 droplets.
To this aim, at each time, we collect several images, in different points of the sample (at least 10) .

\subsection{Scaling arguments for size distribution}
\label{Sec:scaling of size distribution}
We revisit here the main aspects of the theory of breath figures \cite{fam88,fam89}.
Specifically, we address predictions of the time-asymptotic scaling of the droplet number density $n(s,t)$,
and its bearing to the asymptotic power-law decay of the droplet number and the porosity.
Let $r$ be the radius of the circular area covered by the droplet and let the `size' $s=r^3$ be a proxy for the
droplet volume. The size is related to the volume by a constant factor that depends of the wetting angle of the droplets,
and hence on the surface properties.

By applying the Buckingham-Pi theorem \cite{bar03_BOOK}, one can write
\beq
n(s,t) = \Sigma^{-\theta} f(x,y) \, ,
\label{Eq:Buckingham theorem}
\eeq
where $f(x,y)$ is a non-dimensional scaling function. Such a function depends on the dimensionless ratios $x = s/\Sigma(t)$ and $y = s/s_0$,
where $s_0$ is a constant characterizing the smallest droplets in the system and $\Sigma(t)$ the maximum droplet size at
time $t$. By purely dimensional considerations, the exponent $\theta$ must be $\theta = (D+d)/D$ \cite{fam89}, where
$D$ is the dimensionality of the droplets, and $d$ is the dimensionality of the substrate.
In our case, for three-dimensional droplets ($D=3$), on a two-dimensional substrate ($d=2$), we have $\theta = 5/3$.

The theory for breath figures \cite{fam88,fam89} asserts that
the droplet arrangement in the late-time regime is self-similar and features a scaling range.
Moreover, the tails of the distribution, encompassing the smallest and the largest droplets, lie outside
of the scaling range \cite{fam89,bla10}.
Hence, the droplet number density can be expressed as
\beq
n(s,t) =
 K \; \Sigma^{-\theta} \biggl[\frac{s}{\Sigma(t)}\biggr]^{-\tau}
\; \hat{f}\biggl(\frac{s}{\Sigma(t)}\biggr) \; \hat{g}\biggl(\frac{s}{s_0}\biggr)\, ,
\label{Eq:n(s,t)}
\eeq
where $\hat{f}(s/\Sigma(t))$ and $\hat{g}(s/s_0)$ are the so-called 'cutoff functions', describing
the large and small droplets respectively, \ie the tails of the distribution, while $K$
is a constant, chosen in such a way that $\hat{f}(x) = 1$ for small arguments $x$, and $\hat{g}(y) = 1$ for large $y$.
Here, the polydispersity exponent $\tau$ must take a value $0 < \tau < \theta$ to cope with a finite droplet volume
and droplet number at all times~\cite{kol89}. The scaling range of $n(s,t)$ amounts to the interval of $s/\Sigma(t)$
where both cutoff functions, $\hat{f}(x)$ and $\hat{g}(y)$, are constant. It increases over time, since $\Sigma(t)$
grows in time, while $s_0$ is a constant.

\subsection{Scaling arguments for droplet growth exponent}
\label{Sec:largest droplet size}
The radius of the largest droplet in the system, $R$, is expected to grow in time as $R\sim t^\nu$, where $\nu$ is
a constant. Different values have been reported in the literature for the exponent $\nu$
\cite{bey86,per87,vio88,bri91,bri91a,bla12}, depending on the mechanism governing the individual droplets growth
(e.g. surface or bulk diffusion of the vapour, heat dissipation, etc.). Typical values of $\nu$ range from $1/9$ to $1/3$,
in the monodisperse non-coalescing growth phase \cite{bey86,bri91,bri91a}, and from $1/3$ to $1$ in the polydisperse
growth phase \cite{bri91a,bla12}.
In line with previous observations, our data show the values $\nu = 1/3$ in the initial nucleation phase (called (i)
in \Fig{stages}), and $\nu = 1$ in the late-time regime (called (iv) in \Fig{stages}).
Such values can be explained by purely geometrical considerations, under the following assumptions:
\\
1. All droplets take the shape of spherical caps, such that their volume $V$ is proportional to the cube of the radius
$R$ of the wetted area. In particular, $V = \alpha \Sigma$, where $\Sigma = R^3$ is the size of the droplet, conserved
when two droplets merge and $\alpha = (\pi/3) (2+\cos\theta_c)(1-\cos\theta_c)^2/(\sin\theta_c)^3$, with $\theta_c$ the contact angle,
 measured through the liquid phase. Hence for hemispherical droplets, $\theta_c = \pi/2$ and $\alpha = 2\pi/3$.
\\
2. The water flux $\Phi$ (water volume per unit area per unit time) is constant in time and impinges uniformly on the surface.
\\
3. In growth regime (i) there is a constant number $N_0$ of droplets that are roughly monodisperse and equally spaced
such that each droplet covers an area $A_0$. The water flux impinging on the surface is entirely and uniformly distributed between
the droplets, due to surface diffusion. Hence, in the initial phase each droplet collects the flux impinging on a constant area
$A_0$ such that
\begin{align}
  \frac{\rmd (\alpha s)}{\rmd t} &= A_0 \, \Phi = \text{const}
  \\
  &\quad\Rightarrow\quad
    \Sigma(t) = s_0 +  \frac{A_0 \, \Phi}{\alpha} \, (t-t_0)
  \\
  &\quad\Rightarrow\quad
    R(t) = \biggl[  s_0 +  \frac{A_0 \, \Phi}{\alpha} \, (t-t_0) \biggr]^{1/3}
\end{align}
Here $s_0$ and $t_0$ denote the initial droplet size and the initial time, respectively. The droplet size grows linear in time,
and its radius with a $1/3$ power law.
\\
4. In the late stage, self-similar growth regime (iv) the surface is densely covered by droplets such that direct water deposition on the droplets
dominates and each droplet collects the flux impinging on an area proportional to the surface that it covers,
\begin{align}
  \frac{\rmd (\alpha s)}{\rmd t}
  &= \pi \, R^2 \; \Phi
  = \pi \: s^{2/3} \; \Phi
  \\
  &\quad\Rightarrow\quad
    \Sigma(t)
    = \biggl[ s_0^{1/3} + \frac{\pi\Phi}{3\alpha}( t-t_0 ) \biggr]^3
    \label{Eq:Sigma(t)}
  \\
  &\quad\Rightarrow\quad
    R(t) = s_0^{1/3} + \frac{\pi\Phi}{3\alpha}( t-t_0 )
    \label{Eq:Rmax(t)}
\end{align}
The droplet's volume grows asymptotically as $t^3$ while the radius grows linearly in time with a velocity $\pi\Phi/(3\alpha)$, equal to $\Phi/2$ for hemispherical droplets.
Hence, in the late-regime we have $R\sim t^\nu$ and $\Sigma \sim t^{3\nu}$, with $\nu = 1$.

\subsection{Time decay of the porosity}
\label{Sec:derivation p(t)}
In order to determine the time dependence of the porosity, we express the fraction $a$ of the surface area covered by droplets as
\beq
  a(t) =  \int_0^\infty \; C s^{d/D} \; n(s,t)  \; \mathrm{d}s \,
  \label{Eq:a(t)}
\eeq
where $C = \pi$.
We then employ \Eq{n(s,t)} to derive
\begin{align}
  a(t) &=  K' \; \int_0^\infty \; s^{d/D} \; \Sigma^{-\theta} \; \left( \frac{s}{\Sigma} \right)^{-\tau} \;
    \hat{f}\biggl(\frac{s}{\Sigma(t)}\biggr) \; \hat{g}\biggl(\frac{s}{s_0}\biggr)\; \mathrm{d}s
  \\
  &= K' \; \int_{s_p/\Sigma(t)}^{x_p} \;  x^{\theta-\tau-1} \; \mathrm{d}x
  \label{Eq:a(t)}
\end{align}
where $K ' = KC$, $x=s/\Sigma$ and the cutoff functions are taken into account by an informed choice of the constant values $s_p$ and $x_p$ in the integration bounds.
In the long-time limit the lower integration bound, $s_p/\Sigma$, will approach zero and the surface area will be fully covered such that
$a = 1$.
Consequently,
\begin{align}
  1
  =  K' \; \int_{0}^{x_p} \;  x^{\theta-\tau-1} \; \mathrm{d}x
  =  K' \; \frac{ x_p^{\theta-\tau }}{ \theta-\tau } \, ,
\end{align}
from which we derive
\beq
  K' = \frac{ \theta-\tau }{x_p^{\theta-\tau }}
\label{Eq:expression K'}
\eeq
By making use of Eqs.~\ref{Eq:a(t)} and \ref{Eq:expression K'} , the porosity can then be written as
\begin{align}
  p(t)
  &= 1 - a(t)
  =  \left[ \frac{s_p}{ x_p \; \Sigma(t)} \right]^{\theta-\tau}
  \label{Eq:p(t),1}
\end{align}
We substitute \Eq{Rmax(t)}, inside \Eq{p(t),1} and we derive
\beq
p(t) = \left(\frac{s_p}{x_p}\right)^{\theta-\tau} \, R_{max}^{-k} \, ,
\label{Eq:p(Rmax)}
\eeq
where
\beq
k = 3\nu(\theta - \tau)
\label{Eq:exponent k, porosity decay}
\eeq
from which
\beq
p(t) = \left(\frac{27 s_p \alpha^3}{\pi^3 x_p \Phi^3}\right)^{\theta-\tau} \, t^{-k} \, .
\label{Eq:p(t)}
\eeq
Thus the porosity decays in time as $p\sim t^{-k}$.

For Blackman's prediction $\tau^* = 19/12$ \cite{bla00} in combination with the values $\nu = 1$ from geometrical considerations
and $\theta = 5/3$ from dimensional analysis, we find that $p(t) \sim t^{-1/4}$.
Note that the observed decay of the porosity implies that $\tau < \theta$.
We conclude the analysis by taking a closer look at the prefactor of the power law.
The ratio in the brackets amounts to the width of the scaling range,
with cutoffs at the non-dimensional sizes $s_p / \Sigma(t)$ and $x_p$ in Fig.~\ref{Fig:prob density}.
For a given size $\Sigma(t)$ of the largest droplet, the width of the scaling range for different systems
is expected to differ mostly due to differences in the lower cutoff function \cite{bla00}.
Due to the exponent $\theta-\tau = 1/12$, even a difference of two decades for
vastly different materials, would imply that the prefactor of the power law varies by at most $30\%$.

\subsection{Time decay of the number of droplets}
\label{sec:derivation N(t)}
\begin{figure}
\includegraphics[width=0.48\textwidth]{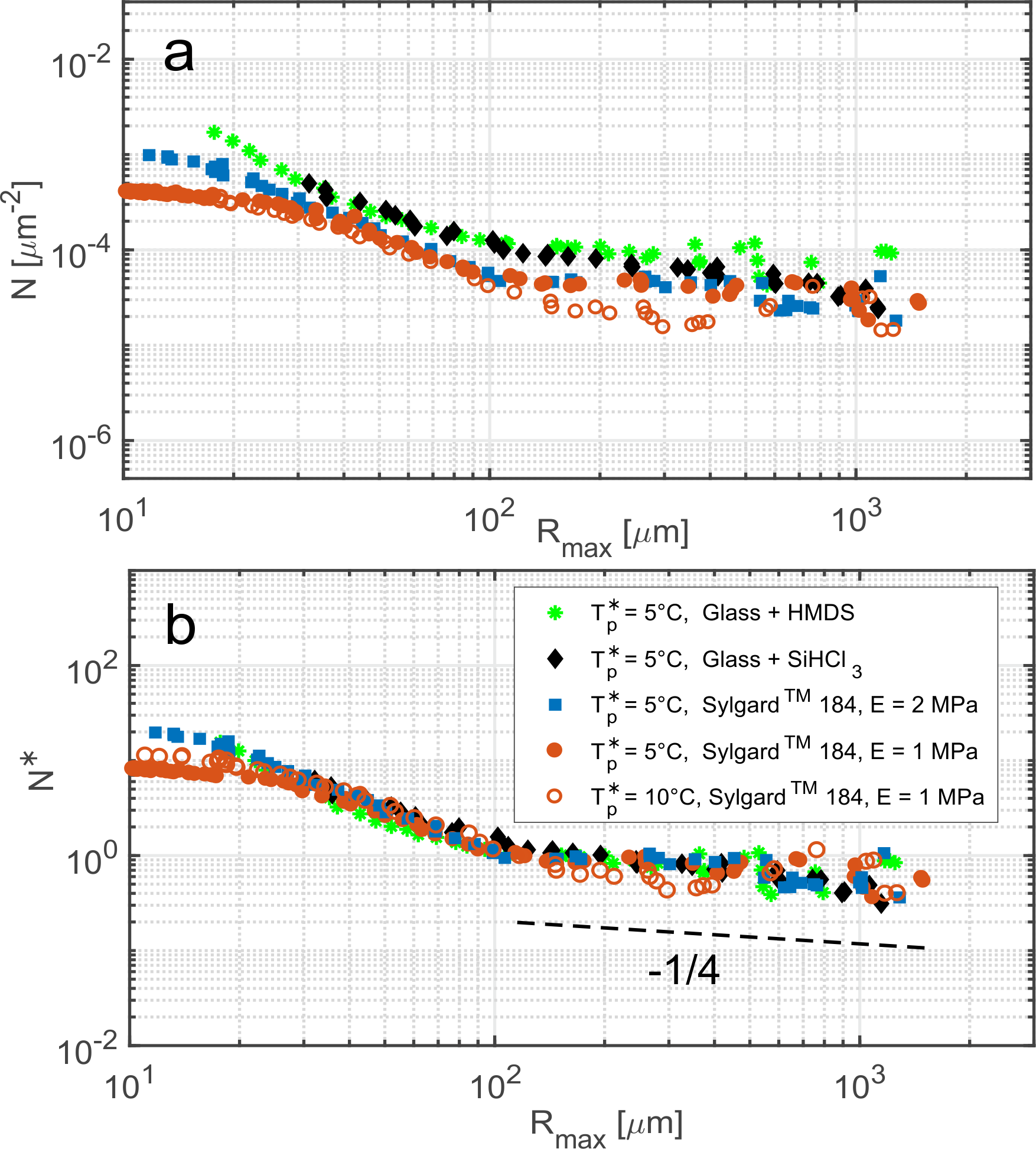}
\caption{\label{fig:Nt-Rmax}
  Evolution of the number of droplet $N$ as a function of the largest radius $R_{max}$ in non-rescaled (a) and rescaled form (b) with $N^*$ provided in Eq.~\ref{Eq:N*}.}
\label{Fig:N(Rmax) diff rescaled only vertically}
\end{figure}

In view of \Eq{n(s,t)} one can write the number of droplets per unit area of substrate, $N(t)$, as
\begin{align}
  N(t) &= \int_0^\infty n(s,t) \: \mathrm{d}s
  \\
       &= K \;  \int_0^\infty \; \Sigma^{-\theta} \; \left( \frac{s}{\Sigma} \right)^{-\tau} \;
         \hat{f}\biggl(\frac{s}{\Sigma(t)}\biggr) \; \hat{g}\biggl(\frac{s}{s_0}\biggr)\; \mathrm{d}s
  \\
       &= K \; \Sigma^{-d/D} \; \int_{s_N/\Sigma(t)}^{x_N} \;  x^{\theta-\tau-1} \; \mathrm{d}x
  \\
  &=  \frac{K}{1-\tau} \; \Sigma^{-d/D} \;
  \left[ x_N^{1-\tau} - \left( \frac{s_N}{\Sigma(t)} \right)^{1-\tau} \right] \, .
  \label{Eq:N(t),1}
\end{align}
Here $s_N$ and $x_N$ in the integration bounds are constants chosen in such a way to keep into account the
cutoff functions. For $\tau<1$ the second term in the square bracket in \Eq{N(t),1} is sub-dominant, and the total droplet number is obtained
by dividing the total area $A_{\text{tot}}$ of the surface by the area $\Sigma^{d/D}$ covered by a typical large droplet.
In view of Fig.~\ref{Fig:prob density}, this is at variance with the finding that small droplets dominate the droplet size distribution in the late-regime. Hence, in the late regime, $\tau > 1$ and
\beq
  N(t) \sim \Sigma^{\tau - 1+d/D} = \Sigma^{\tau - \theta} \sim t^{-3 \; (\theta-\tau)} \, .
\label{Eq:N(t),2}
\eeq
With $\theta = 1 - d/D$ and $\Sigma = R_{\text{max}}^3$ one obtains that
\begin{align}
  N^* = \frac{ N(t) }{ K \, s_N^{1-\tau} } = R_{\text{max}}^{-k} \, .
  \label{Eq:N*}
\end{align}
Hence, in dominant order the droplet number $N(t)$ shows the same power-law decay as the porosity, namely $N \sim t ^{-k}$.
Its dependence as function of $R_{\text{max}}$ is provided in Fig.~\ref{fig:Nt-Rmax}.
However, the data analysis is complicated in this case because the power law suffers from a cross over, even in the large-time asymptotic regime (see \Eq{N(t),1}). Therefore, whenever possible, the analysis of the power law time decay
of the porosity should be preferred as a way to assess scaling in the late regime, respect to the analysis of the number
of droplets.

\subsection{Kinetic Monte Carlo simulations}
\label{Sec:simulations}
The numerical simulations are performed with a lattice Kinetic Monte Carlo algorithm \cite{gil76,bat08,jan12_BOOK} implemented in Python using object-oriented programming.
The code used in this work is a direct adaptation of the code developed in \cite{soe18} for the description of atoms aggregating in 3D clusters over surfaces.
The simulations are performed in a non-dimensional fashion. In order to present the results in dimensional form, the appropriate length and time units are chosen in such
a way that the numerical late time evolution of the largest droplet matches the experimental one (Fig.~\ref{Fig:Rmax, simulations}).

The computational domain consists of a square lattice of $N_c = 1200\times1200$ cells (the 'sites') with periodic boundary conditions.
Each site can be 'empty' or 'occupied', \ie completely filled with water. Each droplet consists of a set of contiguous occupied sites, symmetrically arranged around
the droplet's center. Therefore, each droplets is a circle approximated by its staggered version, in perfect analogy with the pixel resolution of the experimental optical measurements.
Initially, no droplets are present and all sites are empty. As time progresses, water is deposited and the droplets start to nucleate, grow and merge
with each other. Time progression happens in a discrete fashion, assuming that, at each time $t_j$ an instantaneous 'event' (deposition of a discrete amount of water)
occurs. The event is then followed by a waiting time, \ie a time step, $\Delta t_j$, when nothing happens. Thus, the next time instant will be $t_{j+1} = t_j+\Delta t_j$.
Note that the time intervals $\Delta t_j$ are not constant, but randomly chosen from a Poisson distribution, since the events are considered as rare and
independent from each other \cite{gil76,bat08,jan12_BOOK}. The average of such a distribution is $1/k_{tot}$, where $k_{tot} = \hat{\Phi} A_{tot}$
with $A_{tot}$ the total area of the substrate, namely the total number of sites and
$\hat{\Phi}$ the number of events (falling droplets) per unit time per unit surface, hence $k_{tot}$ represents the total frequency at which the events
take place on average, \ie the number of events per unit time. The flux $\Phi$, i.e. the water volume deposited per unit area, can be determined as $\Phi = \hat{\Phi}V_0$,
where $V_0$ is a constant representing the volume deposited in a single event. However, changing $\Phi$ is equivalent to a mere time rescaling \cite{soe18} and does not
affect the critical exponents. Therefore one can take both $\hat{\Phi}$ and $V_0$ equal to unity and choose the time and length units to achieve dimensional matching with both
the size of the smallest droplet detected in the experiments and the experimental late time evolution of the largest droplet.

\begin{figure}
\includegraphics[width=0.95\columnwidth]{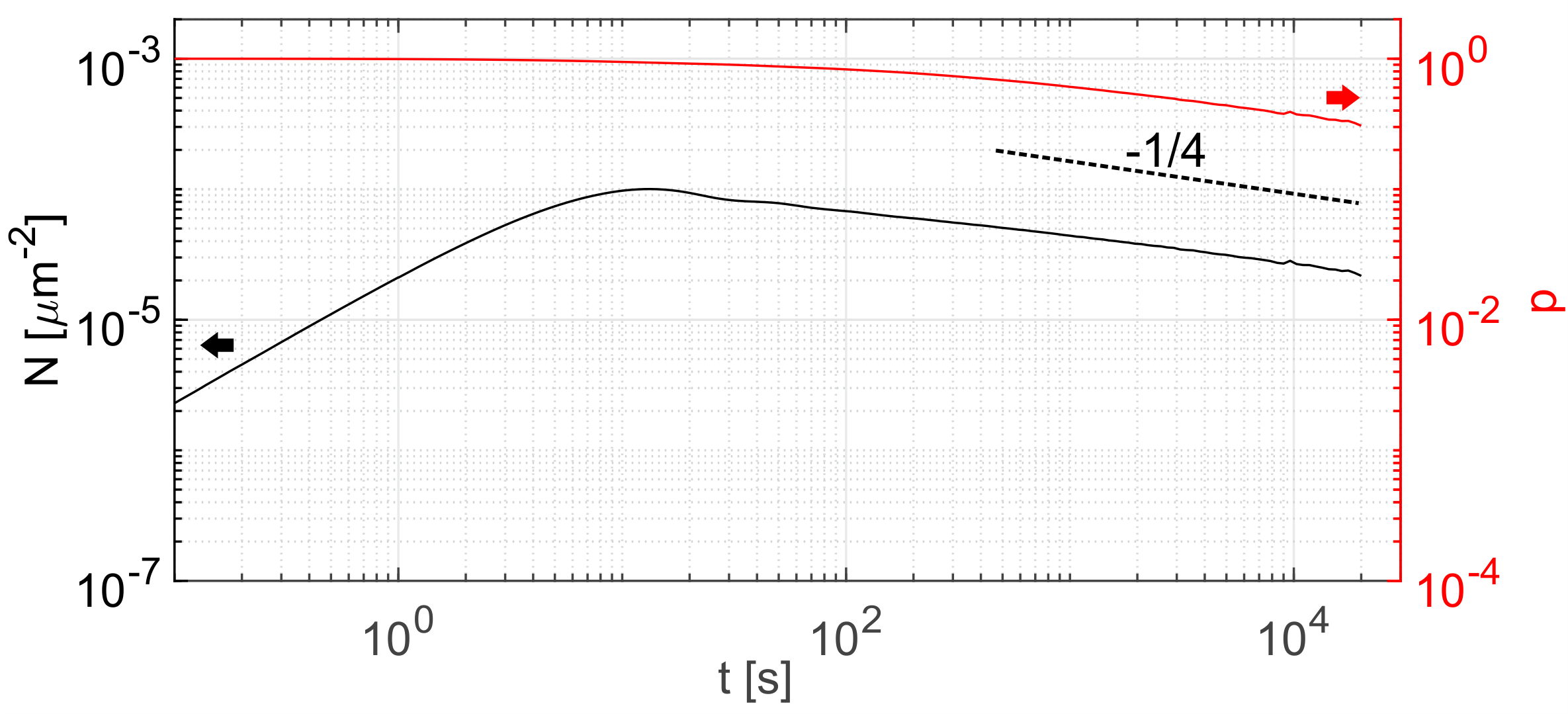}
\caption{Numerical results for time evolution of porosity (grey, red online, right vertical axis) and number of droplets per unit area (black, left vertical axis),
averaged over 11 simulations. The dashed line represent the theoretical prediction for the time decay of both the porosity and the number of droplets.}
\label{Fig:p(t),N(t) simulations}
\end{figure}

Once the time step to advance the system has been decided, the algorithm chooses the type of event that takes place.
In particular, two types of events can occur: the nucleation of a new droplet on an empty site or the growth of an existing
droplet due to direct water deposition. The type of event is randomly determined, at each time, based on the probability that either growth or nucleation occurs.
Such probabilities are $N_{c,occ}/N_c$ for droplet growth, and $(N_c-N_{c,occ})/N_c$ for droplet nucleation, where $N_{c,occ}$ is the
total number of occupied sites. The volume of deposited water at each time step is constant, analogous to the deposition of a droplet.
At this point, once the type of event (droplet growth or nucleation) has been decided, the exact position
where the water volume will fall is determined with a third random number, based on the probabilities associated to each existing droplet and nucleation site.
In particular, if the event is a droplet's growth, each existing droplet $i$ will have a probability $N_{i}/N_{c,occ}$ to grow, where $N_{i}$ is the number of sites
occupied by the droplet $i$. If the event is a nucleation, each free site will entail a probability $1/(N_c - N_{c,occ})$ to host the new droplet.
Whenever one of these events leads to overlapping between droplets, \ie occupation of the same site by more than one droplet, all the involved droplets are merged.
The merging of droplets is assumed to be instantaneous and preserves the droplets' volume. A droplet resulting from a merging event has the same volume as the sum of the
merging droplets and the center located at the center of mass of the system formed by the two merging droplets. The process is sequentially repeated until no more overlapping between
droplets is present and each site is occupied at most by a single droplet.

\begin{figure}
\includegraphics[width=0.85\columnwidth]{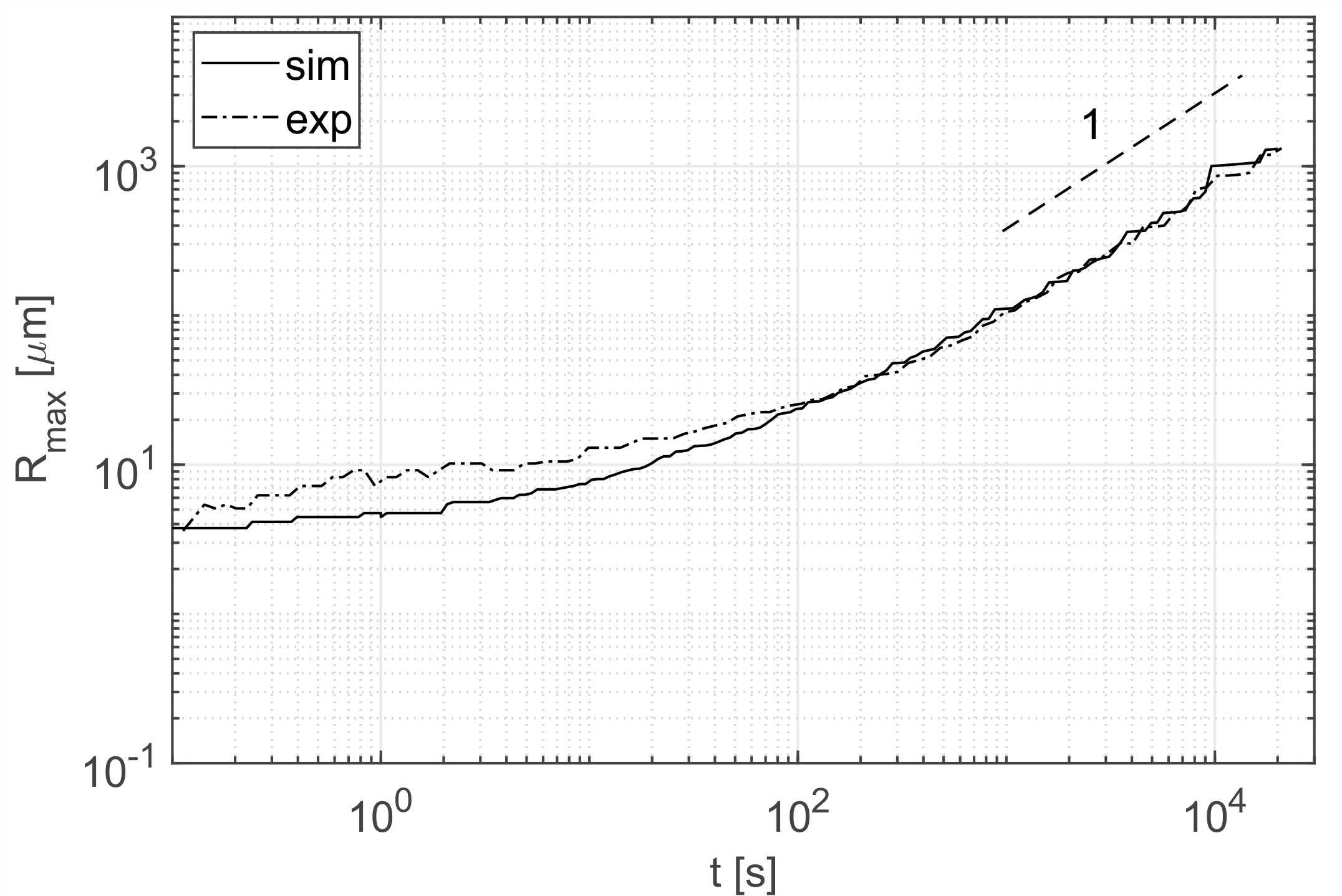}
\caption{Time evolution of the largest droplet in simulations (solid line), dimensionally matched to the experiments presented in Fig.~\ref{Fig:stages} (dash-dotted line).}
\label{Fig:Rmax, simulations}
\end{figure}

\newpage

\begin{figure*}
\includegraphics[width=0.7\textwidth]{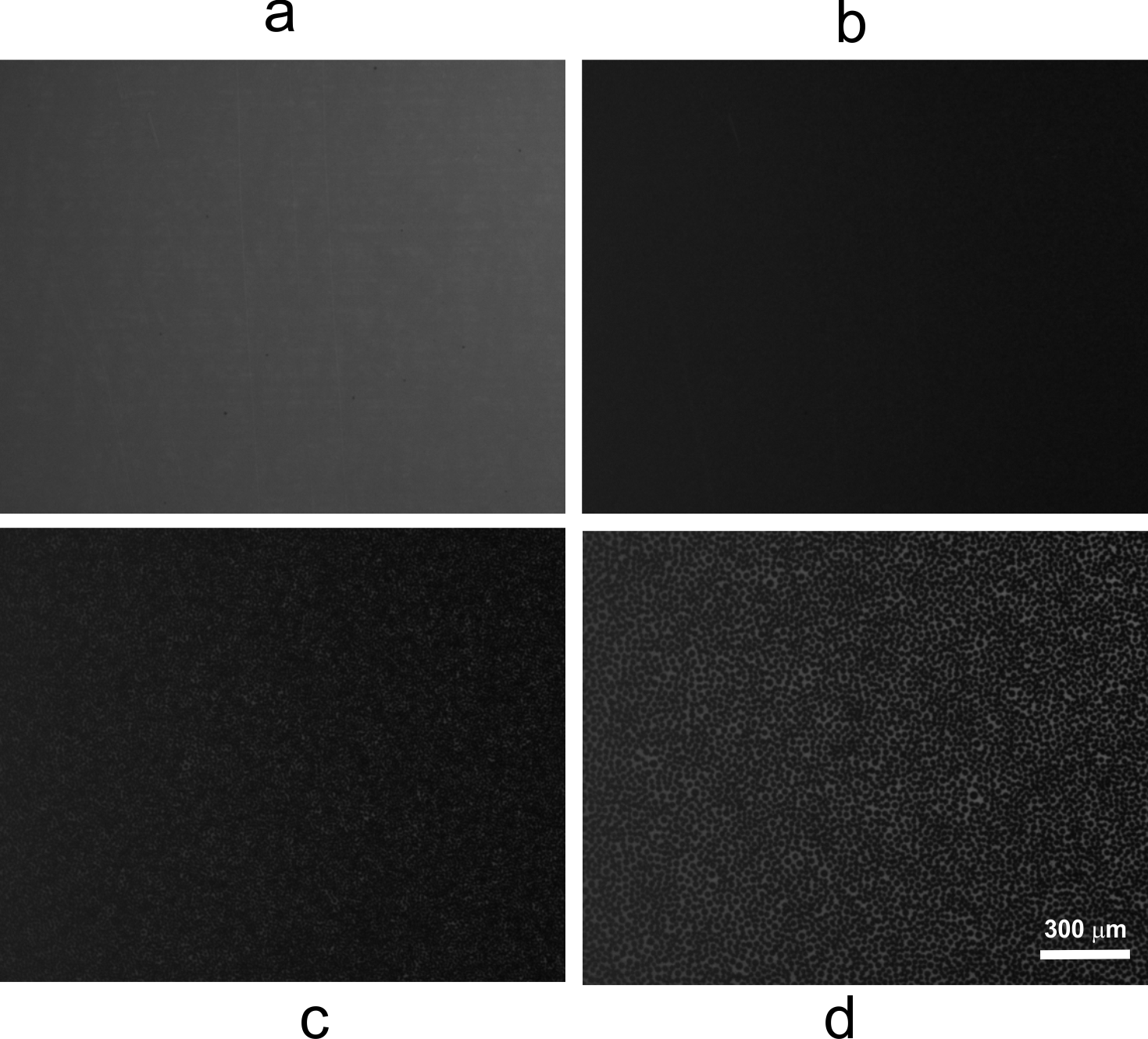}
\caption{Initial phases of water deposition on fluor-silanized glass: at the beginning of the experiment (a), after 1 sec (surface completely covered by water,
possibly in the form of droplets below resolution)(b), 5 sec (c), 10 sec (d).}
\label{Fig:glass, nucleation}
\end{figure*}

\begin{figure*}
\includegraphics[width=0.6\textwidth]{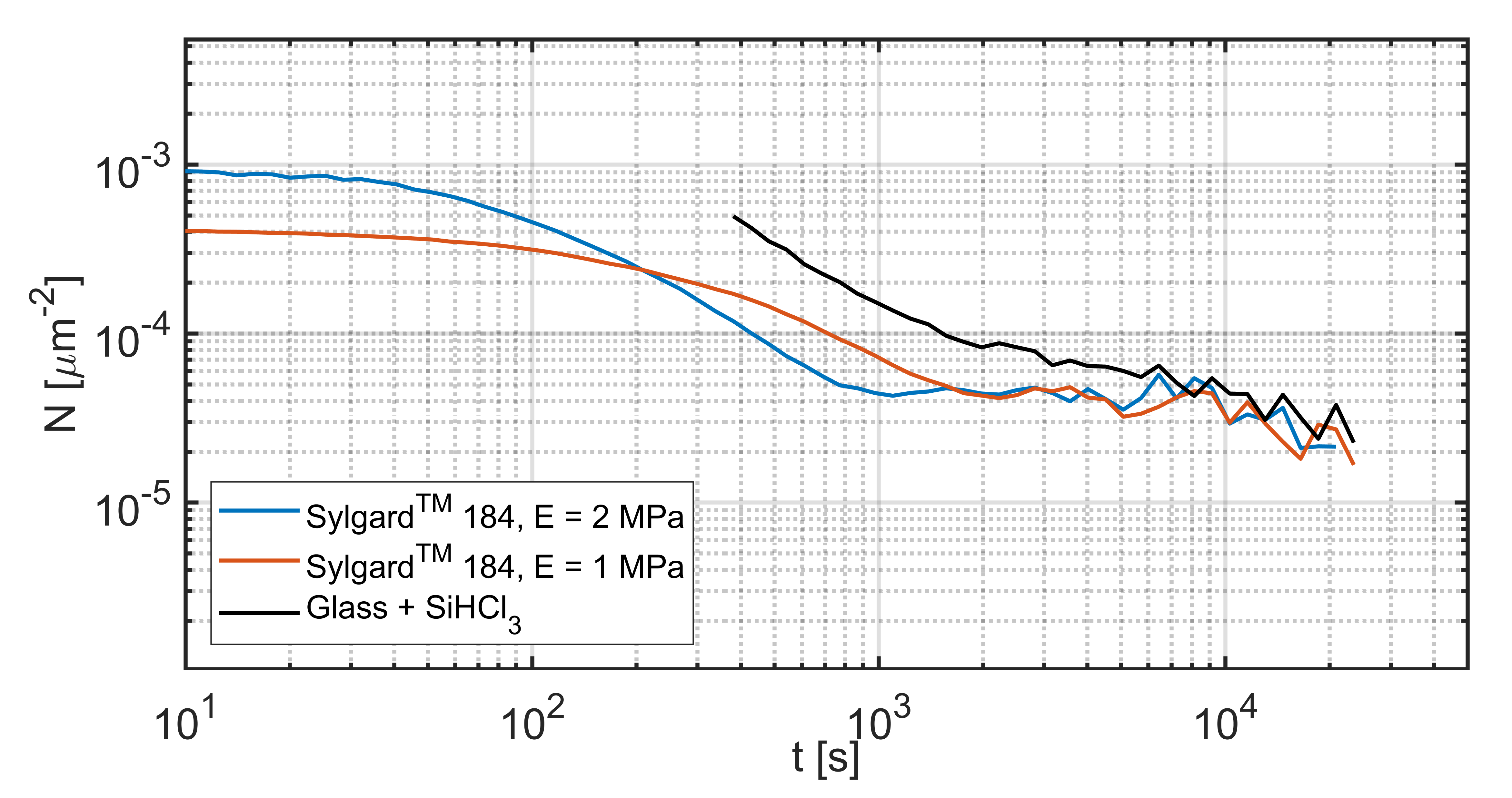}
\caption{Time evolution of the number of droplet $N$ on different surfaces with plate temperature $T_p^*=5\degree$~C, for the same conditions displayed in \Fig{porosity diff}.a.}
\label{Fig:N(t) diff}
\end{figure*}

\begin{figure*}
\includegraphics[width=0.6\textwidth]{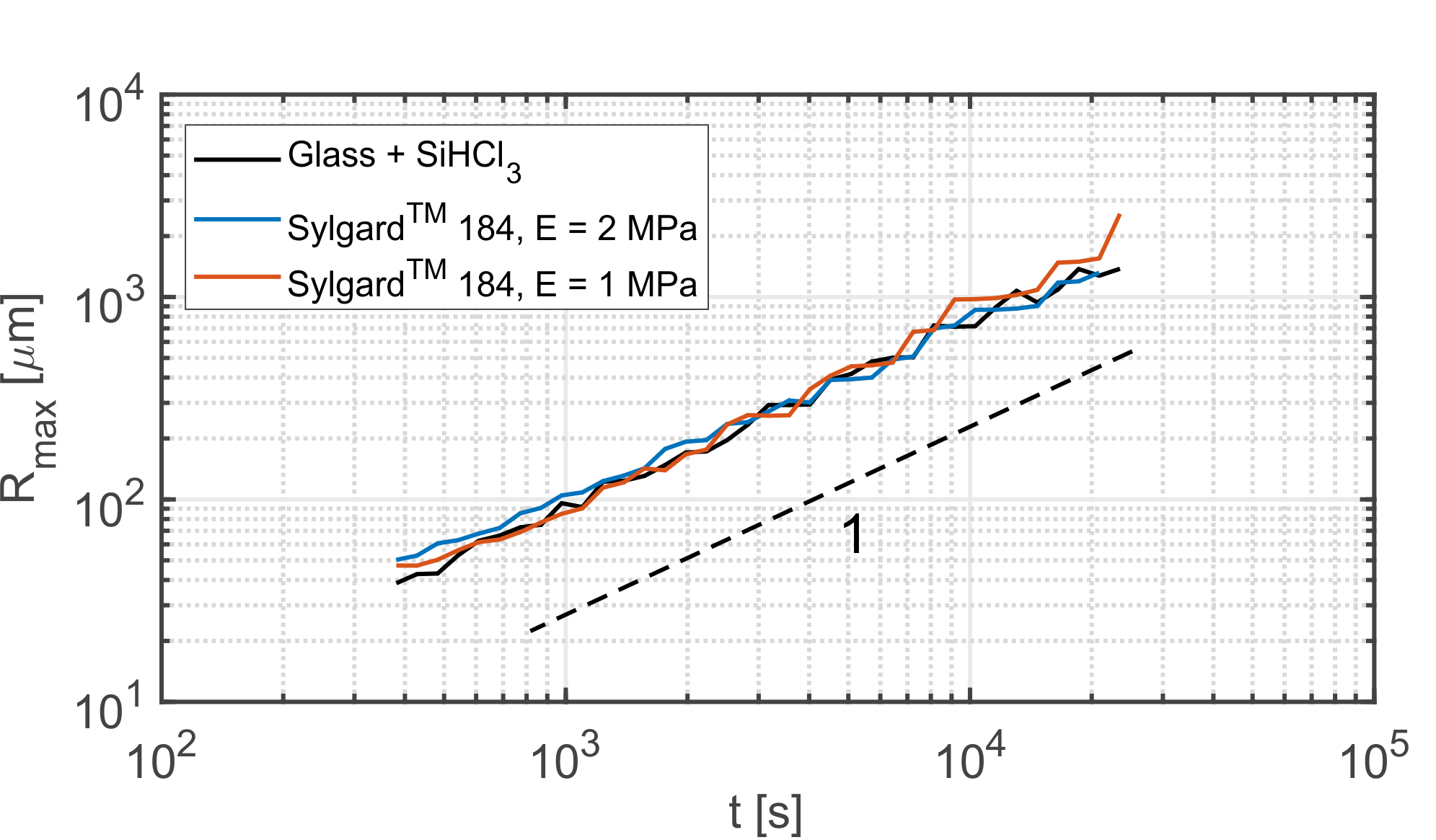}
\caption{Time evolution of the largest radius $R_{max}$ for different rigid surfaces with plate temperature $T_p^*=5\degree$~C, for the same conditions displayed in \Fig{porosity diff}.a.}
\label{Fig:Rmax(t) diff before rescaling}
\end{figure*}

\end{document}